\documentclass[twoside]{article}
\usepackage{fleqn,espcrc2}

\usepackage{graphicx}

\newcommand{\e}{\mbox{e}}

\newcommand{\AmS}{{\protect\the\textfont2
  A\kern-.1667em\lower.5ex\hbox{M}\kern-.125emS}}

\title{Discrete Lorentzian Quantum Gravity}

\author{R. Loll\address{Albert-Einstein-Institut,
Max-Planck-Institut f\"{u}r Gravitationsphysik,\\
Am M\"uhlenberg 1, D-14476 Golm, Germany}
}

\begin{document}

\begin{abstract}
Just as for non-abelian gauge theories at strong coupling, 
discrete lattice methods 
are a natural tool in the study of non-perturbative quantum
gravity. They have to reflect the fact that the geometric
degrees of freedom are dynamical, 
and that therefore also the lattice theory must 
be formulated in a background-independent way.
After summarizing the status quo of discrete covariant lattice
models for four-dimensional quantum gravity, I describe a
new class of discrete gravity models whose starting point is
a path integral over Lorentzian (rather than Euclidean) space-time
geometries. 
A number of interesting and unexpected results that have been
obtained for these dynamically triangulated models in 
two and three dimensions
make discrete {\it Lorentzian} gravity a promising candidate 
for a non-trivial theory of quantum gravity.

\vspace{1pc}
\end{abstract}

\maketitle

\section{QUANTUM GRAVITY}

For the purposes of this presentation, 
{\it quantum gravity} will be defined as
the non-perturbative quantization of the classical theory of
general relativity, with and without the inclusion of other matter
fields and interactions. Judging from our current knowledge of the 
fundamental
laws of physics, it seems highly likely that at sufficiently large
energies also {\it gravitational} interactions should be governed by 
quantum rather than by classical equations of motion. 
Quantum gravity -- whose theoretical formulation is still elusive --
should include a consistent description of local quantum phenomena in the
presence of strong gravitational fields. 

It has been known for a long time that {\it perturbative} quantum 
gravity, based on a decomposition
\begin{equation}
g_{\mu\nu}(x)=\eta_{\mu\nu}+\sqrt{32\pi G}\ h_{\mu\nu},
\label{gdecomp}
\end{equation}
of the Lorentzian space-time metric $g_{\mu\nu}$ into the flat
Minkowski metric $\eta_{\mu\nu}$ and a linear perturbation 
$h_{\mu\nu}$ (representing the degrees of freedom of a massless
spin-2 graviton), leads to a non-renormalizable field theory.
Although this does not preclude the use of the perturbation series
as an effective description of quantum gravity in the presence of
an energy cut-off, it cannot serve as the definition of a
fundamental theory. 

The ensuing need to quantize gravity {\it 
non-perturbatively} is not confined to field theory,
but persists in string-theoretic formulations 
(where it is an unsolved problem as well).
Imagine trying to obtain a quantum state representing a
4d Schwarzschild black hole with metric
\begin{equation}
g_{\mu\nu}=-(1-\frac{GM}{r})dt^{2}+(1-\frac{GM}{r})^{-1}
dr^{2}+r^{2}d\Omega^{2}
\label{schwarzschild}
\end{equation}
by superposing gravitonic excitations within string theory. 
However, because of the proportionality $G\sim g_{\rm str}^{2}$ for
Newton's constant $G$ \cite{pol}, this
involves arbitrary powers of the string coupling $g_{\rm str}$, and
is therefore an intrinsically non-perturbative construction.

\subsection{How do we quantize gravity non-perturbatively?}

The great success of lattice models in describing non-perturbative
properties of QCD has for a long time been a motivation for 
applying discrete methods also in quantum gravity. 
I will be reporting on the status quo of path-integral 
(``covariant'') lattice models for quantum gravity, and on
how to make them more {\it Lorentzian}. However, it should
be pointed out that there are other ways of tackling the problem,
most notably, in a canonical
continuum approach based on a gauge-theoretic formulation of gravity
in four dimensions \cite{rovthie}. 

Non-perturbative formulations of gravity usually involve suitable
versions of the space $\cal M$ of all 
4d space-time geometries 
(of space-time metrics modulo diffeomorphisms), and all
lattice models start from a discretized version of this
classical configuration space. Note that giving up the
notion of a preferred flat background metric $\eta_{\mu\nu}$ also
means that the Poincar\'e group will {\it not} appear as a fundamental
symmetry group of the base space $M$. 
This has nothing to do with the loss of the smooth
manifold structure during discretization, but is a typical feature
of all non-perturbative, background-independent formulations.

Among the important (and difficult) questions one may ask in discrete 
quantum gravity are the following: \\
(i) how do we construct quantum gravity lattice models 
that are well-defined, convergent statistical systems
at finite volume? How is the absence of a preferred background metric
reflected in the lattice construction? \\
(ii) Do these models lead to
interesting continuum theories in the infinite-volume limit? What
are their physical excitations? \\
(iii) What is the (quantum) geometry of the ground state of the
theory? Do we recover semi-classical geometries in a suitable limit?

\subsection{Where do we stand?}

The three most popular approaches to discrete quantum gravity in four
dimensions are: \\
{\it Covariant Gauge Approaches}, based on gauge-theoretic 
first-order formulations of gravity with connections and vierbeins
(mostly on regular cubic lattices); they were studied intensely 
for about ten years, starting in the late 70s \cite{cga}. More recently,
discrete gauge-theoretic formulations have seen a revival in the
context of so-called spin foam models for gravity, based on ideas
coming from topological quantum field theory \cite{foam}.\\
{\it Quantum Regge calculus} is based on the
second-order form of gravity in terms of metric fields. 
Space-time is approximated by
a simplicial complex, and the sum over all geometries takes the
form of a sum over all possible lengths of the edges of this complex.
This approach to quantum gravity originated in the mid-80s and is
still being pursued \cite{qrc}.\\
The method of {\it Dynamical Triangulations} is 
a more recent variant of the quantum Regge calculus 
program, where all edge lengths are frozen in, and the state sum is
taken over all possible manifold-gluings of a set of
equilateral simplicial building blocks. Although this is the most
recent of the three approaches 
(started in the early 90s \cite{aj,am}), the 
number of research papers written in this area is by far the largest. 
Interest in this formulation was largely propelled by its success in 
reproducing results from continuum Liouville gravity in
$d=2$. --
I will not dwell on a detailed description of the individual 
achievements of these approaches, since this has been done both
in previous overview talks at lattice conferences, and in a recent
review article on the subject \cite{loll1}. 

What has been learned
from these investigations? Maybe not surprisingly,
analytic results in 4d have proved hard to come by, although
qualitative estimates of the partition function in certain phases
can sometimes be made. Efficient numerical methods for models of
fluctuating geometry and/or lattices have been developped and refined.
The phase structures of all of the models described above have been
investigated and their phases characterized in geometric terms,
showing some surprising similarities across the various formulations
\cite{loll1}. However, it is probably fair to say that {\it in spite of
occasional claims to the contrary, there is so far no convincing 
evidence of a second-order phase transition} in these models,
which is usually taken as a necessary condition for the existence of
an underlying continuum theory with propagating field degrees of
freedom.

\subsection{What to do?}

Various strategies have been tried to improve on this somewhat 
unsatisfactory state of affairs, usually by modifying the measure
of the quantum theory, or by adding appropriate matter fields.
I refer to last year's review talk by Krzywicki \cite{krzy} 
for a more detailed
account of these attempts. The most recent development in this
area concerns the addition of non-compact U(1)-gauge fields in
the dynamical triangulations approach. The phase structure of these
models does indeed appear changed with respect to the pure-gravity
models \cite{bilke}, but more recently it has been understood that
this is a ``spurious'' effect, and completely equivalent to a
rescaling of the measure by factors of the determinant of the metric 
\cite{amb1}. Conflicting numerical results on this issue
have been reported at this conference \cite{horata}, so this may still
hold out some hope for more interesting findings in this matter-coupled
model. However, the general consensus seems to be that more radical
changes of the discrete models are required to alter these negative 
results.

\subsection{Absence of a Wick rotation}

Unfortunately, until recently it seemed that we had run out of
natural, geometrically motivated ways of modifying the discrete
gravity models. There is another reason for why the situation may
be yet more serious: even if some of these
quantization attempts {\it had} been successful, 
their connection with the physical, Lorentzian theory
of quantum gravity would have remained unclear, because they all 
work with configuration
spaces of (positive-definite) {\it Euclidean} geometries. 
There are of course good technical reasons for making the
substitution
\begin{equation}
\int\limits_{\frac{{\rm Lor}(M)}{{\rm Diff}(M)}} 
[{\cal D}g^{\rm Lor}_{\mu\nu}]
\e^{i S[g^{\rm Lor}]}\rightarrow 
\int\limits_{\frac{{\rm Eu}(M)}{{\rm Diff}(M)}} [{\cal D}g^{\rm Eu}_{\mu\nu}]
\e^{- S[g^{\rm Eu}]},
\label{subst}
\end{equation}
since state sums over complex phase factors usually diverge.
Well-motivated though it may be, it should be pointed out that
in a non-perturbative context, (\ref{subst}) is an {\it ad hoc}
substitution. Although (\ref{subst}) may look suggestive, 
unlike in ordinary field theory on a Minkowski
background, there is no {\it a priori} concept of a ``Wick rotation''
in a theory with a dynamical metric. The problem lies in the fact
that we must Wick-rotate {\it all} metrics, but that almost all
metrics $g_{\mu\nu}$ (unless they have special symmetries,
for example, time-like Killing vectors) 
have no geometrically distinguished notion of
``time''. (Recall that $t$ in general relativity denotes simply
``coordinate time'', and can be changed by a diffeomorphism without
affecting physical results.) However, a prescription like
$t\mapsto\tau =-it$ is certainly 
{\it not} diffeomorphism-invariant (think of a
simple coordinate transformation like $t\mapsto t^2$, for $t>0$).

Since there is a certain reluctance to recognize this as a {\it
problem}, let me add some further remarks on the issue.
It does of course not follow from the discussion above 
that suitable Wick rotations do {\it not} exist. However, 
they should be
constructed explicitly, and their naturalness and/or uniqueness be shown.
Also, one might hope that once an interesting continuum limit of
{\it Euclidean} quantum gravity is found, a continuation to Lorentzian
signature might be obvious. This is a logical possibility, but it has
not yet been realized.

At any rate, there is no {\it a priori} reason that a theory based 
on a non-perturbative path integral for Riemannian (or Euclidean)
metrics should be related in any simple way to one based on 
Lorentzian metrics. Interpreted positively, this observation
may offer us a way out of the current impasse in finding
physically interesting quantum gravity models. Our failed attempts to
quantize could be closely related to the fact that the Lorentzian
nature of gravity is not appropriately taken into account. 
As I will describe in more detail below, there is now evidence from
discrete Lorentzian models of gravity in two and three space-time 
dimensions supporting this point of view. 

\subsection{Going Lorentzian}

\begin{figure}[h]
\vspace{-0.8cm}
\centerline{\scalebox{0.35}{\rotatebox{0}
{\includegraphics{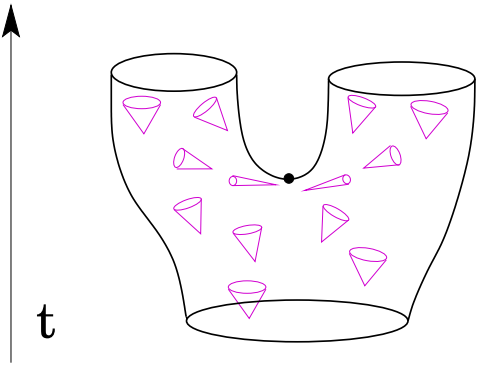}}}}
\vspace{-1.0cm}
\caption[trousers]{At a branching point associated with
a spatial topology change, light-cones get ``squeezed''.}
\label{trousers}
\end{figure}

In the continuum, 
a Lorentzian space-time is usually given in the
form of a metric field tensor $g_{\mu\nu}(x)$ on some manifold
$M$, with signature
(--+++), where $g_{\mu\nu}$ is symmetric and non-degenerate, but not
necessarily diagonal. The associated line element
\begin{equation}
ds^2=g_{\mu\nu}(x) dx^\mu dx^\nu
\end{equation}
may therefore take values $<0$, $=0$ or $>0$, depending on whether the
infinitesimal distance measured on $M$ is time-like, 
light-like or space-like. It follows that the neighbourhood of any point
$p\in M$ has a light-cone structure, where the light-cone consists
of all points $q$ that can be connected to $p$ by curves whose
tangent vectors are everywhere light-like, with analogous prescriptions
for points in- and outside the light-cone. Just as it does in
Minkowski space, this implies a local causal structure: a point $q$
lies to the future (past) of $p$, if there is a future- (past-)oriented
curve from $p$ to $q$ that is nowhere space-like. Otherwise, $p$ and
$q$ are not causally related. In addition, one usually requires 
a well-defined causal structure {\it globally}, to avoid pathologies
such as closed time-like curves. Note also that branching points,
associated with a topology change of spatial slices 
(Fig.\ \ref{trousers}) are not
compatible with a well-defined Lorentzian structure, since 
the light-cones must necessarily degenerate at such points.
-- By contrast, in Euclidean metric spaces there is no distinction 
between space- and time-like directions. 
How can Lorentzian
features be built into a framework of discretized geometries? Regge's
prescription for approximating smooth geometries by piecewise linear
spaces works just as well (and was originally conceived) 
for Lorentzian signature. The work described below may be regarded as
a Lorentzian version of the dynamical triangulations (DT) approach to
quantum gravity. We prefer this method over quantum Regge calculus,
since we are interested in an analytic formulation (which even for
$d<4$ is impossible in Regge calculus, due to the presence of triangle
inequalities) and because the evidence from $d=2$ suggests that
DT deals correctly with the diffeomorphism symmetry of the theory.

In a Lorentzian DT approach, one may expect to have both time- and
space-like edges (and possibly even null-edges). However, a random
assignment of squared edge lengths $l^2=\pm 1$, say, to an 
arbitrary simplicial complex (a ``triangulation'') will in
general not lead to a metric structure of the correct signature and
without closed time-like curves. Our strategy will be to first identify
a large class of well-defined discrete causal triangulations
(without restricting the local curvature degrees of
freedom). 
In order to make the associated partition function convergent, we
will then use a Wick rotation to map each discrete Lorentzian geometry
into a unique Euclidean geometry. After the sum and the continuum
limit have been performed, the propagator is ``rotated back''. 
Our particular choice of the fundamental
building blocks, described below, 
is motivated by a simple form for both the
path integral and the Wick rotation.

\section{THE NEW IDEA}

\begin{figure}[h]
\vspace{-0.5cm}
\centerline{\scalebox{0.35}{\rotatebox{0}
{\includegraphics{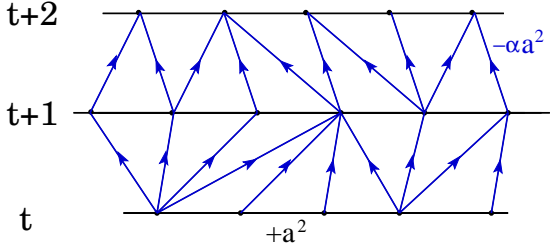}}}}
\vspace{-.8cm}
\caption[2dtriang]{Two layers of a 2d Lorentzian triangulation,
with spatial slices of constant $t$ 
and interpolating future-oriented time-like
links.}
\label{2dtriang}
\end{figure}

Let me now turn to an explicit description of the Lorentzian DT
model, which incorporates a notion of causality and possesses a
``Wick rotation'' \cite{al,ajl1,ajl2}. 
The partition function takes the form of
a sum over causal triangulations $T$ with certain edge length assignments,
\begin{equation}
Z(\lambda,G)=\sum_{{\rm causal}\ T} \frac{1}{C_T} \e^{iS^{\rm Regge}},
\label{part}
\end{equation}
with each contribution weighted by the Regge action (the
simplicial version of the $d$-dimensional Einstein action, including
a cosmological constant $\lambda$) associated
with $T$ and a discrete 
symmetry factor $C_T$. The triangulations appearing
in the sum (\ref{part})
all have a foliated structure, where successive (d--1)-dimensional
spatial slices (realized as equilateral Euclidean triangulations of
squared edge length $l_s^2=+a^2$)
are connected by time-like, future-oriented edges of length-squared 
$l_t^2=-
\alpha a^2$, with $\alpha >0$. This is most easily illustrated in
1+1 space-time dimensions (Fig.\ \ref{2dtriang}), 
where the spatial slices are
simply given by chains of $n$ space-like edges. In addition,
to reflect the causal properties of the continuum geometries these
piece-wise linear spaces are supposed to approximate, we do not
allow for any spatial topology changes. 
For simplicity, we use
compactified and connected spatial slices, yielding a space-time 
topology $R\times S^1$.  

As mentioned earlier, we can now define a unique Wick rotation on 
any discrete Lorentz geometry of this type by substituting all
time-like links with $l_t^2=-\alpha a^2$ by space-like links
with squared edge length $l_t^2=+\alpha a^2$. One can show 
\cite{ajl1,ajl2}
that this analytic continuation in $\alpha$ has the
desired effect of inducing
\begin{equation}
\e^{iS^{\rm Lor}}\;\stackrel{l_t^2\rightarrow -l_t^2}{\longrightarrow}
\;\e^{-S^{\rm Eu}}
\end{equation}
on the weights. It is important to realize that the set of Euclidean
triangulations obtained after the Wick rotation is strictly smaller
than the set of all Euclidean triangulations. It is precisely 
this feature
that leads to a change of universality class of the Lorentzian
models, compared with the Euclidean ones. One could rephrase this
by saying that we have introduced a different {\it measure}, which
however was not chosen {\it ad hoc}, but motivated by physical and
geometric considerations. Obviously, in the end only the solution of 
this model and its physical properties can tell us whether this
ansatz is justified. In this regard, we are encouraged by the results 
obtained so far in dimensions two and three.

What has been shown is that for finite lattice volume in $d=2,3$ and 4, 
the discrete Lorentzian models are completely well-defined, in the
sense that the associated transfer matrices $\hat T$ in the
discrete propagators
\begin{equation}
G({\rm g}_1^{(d-1)},{\rm g}_2^{(d-1)},t)=
\langle {\rm g}_1^{(d-1)} |\hat T^t |{\rm g}_2^{(d-1)}\rangle
\end{equation}
(where g$_i^{(d-1)}$ denotes a discrete spatial geometry)
are bounded and strictly positive. The slice parameter $t$ has a
natural physical interpretation as a discrete proper time, that is,
the time experienced by an idealized set of observers freely falling 
along geodesics perpendicular to surfaces of constant $t$.
Our next task will be to understand
the continuum theories associated with these models. Fortunately,
at least in $d=2$, Lorentzian quantum 
gravity turns out to be exactly soluble.

\section{LORENTZIAN GRAVITY IN D=2}

In two space-time dimensions, the Einstein action for a fixed topology 
reduces to the cosmological term
\begin{equation}
S=\Lambda \int d^{2}x \sqrt{|\det g |}\stackrel{\rm discret.}
{\longrightarrow}
S=a^{2}\lambda N_{2}(T),
\end{equation}
where $N_{2}(T)$ is the number of triangles in the triangulation $T$,
and $\lambda$ is the bare cosmological constant. The most natural
propagator in Lorentzian gravity is a ``two-loop function'', 
describing the transition amplitude between an initial geometry of
length $l_{\rm in}$ and a final geometry of length $l_{\rm out}$ 
(with integer lengths $l_i=1,2,\dots$) 
in $t$ time steps. Its functional form after the Wick rotation 
(and setting $a=1$) becomes
simply \cite{al,alnr}
\begin{eqnarray}
G_{\lambda}(l_{\rm in},l_{\rm out},t)&=&\sum\limits_{{\rm causal\ }T}
\e^{-\lambda N_{2}(T)}\nonumber\\
&=&\sum\limits_{N_{2}}\e^{-\lambda N_{2}}
\sum\limits_{{\rm causal\ }T_{N_{2}}}1.
\end{eqnarray}
The last expression on the right makes it clear that solving the model
is tantamount to solving the combinatorial problem of counting the
number of inequivalent causal triangulations of volume $N_{2}$ and
length $t$, for given lengths $l_{\rm in}$ and $l_{\rm out}$. 
(Similar statements hold in higher dimensions too.)

In two dimensions, this problem can be solved explicitly, and
leads after continuum limit, renormalization and an inverse Wick
transformation to the continuum amplitude
\begin{eqnarray}
G_{\Lambda}(L_{\rm in},L_{\rm out},T)=
\e^{-\coth (i\sqrt{\Lambda}T) 
\sqrt{\Lambda}(L_{\rm in}+L_{\rm out})} &&\nonumber\\
\times\ \frac{\sqrt{\Lambda L_{\rm in}L_{\rm out}}}{\sinh 
(i\sqrt{\Lambda}T)}\ I_{1}
\left( 2\frac{\sqrt{\Lambda L_{\rm in}L_{\rm out}}}{\sinh 
(i\sqrt{\Lambda}T)} \right),&&
\label{contprop}
\end{eqnarray}
where $I_{1}$ is the Bessel function, and $\Lambda$, $L$ and $T$ are
the renormalized counterparts of the bare constant $\lambda$ and 
of $l$ and $t$. 
The theory described by (\ref{contprop}) is unitary,
and its Hamiltonian can be mapped onto
a three-dimensional harmonic oscillator with spin $1/2$ 
and diagonalized explicitly \cite{dkl}. 
Modified versions of 2d Lorentzian gravity (including a higher-order
curvature term, and ``decorations'' by dimers and a restricted class
of baby universes) have also been solved analytically \cite{fgk}, 
leading to similar results.

\subsection{Geometry of the ground state in 2d}

What is the physics described by this model? 
There cannot be much physics to speak of, since classical
gravity in 2d is an empty theory. All one can expect are quantum 
fluctuations at the Planck scale (which happens to coincide with the
cosmological scale, since the theory has only a single length scale).
Nevertheless we can investigate the geometric properties of the
ground state of the quantum theory and compare them with the
Euclidean case. 
\begin{figure}[h]
\vspace{-.6cm}
\centerline{\scalebox{0.6}{\rotatebox{0}
{\includegraphics{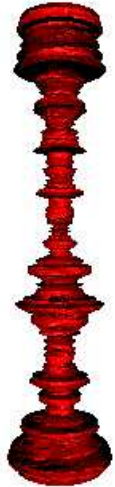}}}}
\vspace{-.8cm}
\caption[puregrav]{A typical 2d Lorentzian space-time, at volume
$N_2=18816$ and total proper time $t=168$.}
\label{puregrav}
\end{figure}
Fig.\ \ref{puregrav} shows a typical 2d Lorentzian space-time, taken
from a Monte Carlo simulation. Observe how the size of the
compactified spatial slices changes as a function of proper 
time (pointing
upwards). These fluctuations are indeed large, and of the same order
as the average spatial length, $\langle \Delta L\rangle\simeq 
\langle L\rangle$. 

\begin{figure}[h]
\vspace{-.6cm}
\centerline{\scalebox{0.5}{\rotatebox{0}
{\includegraphics{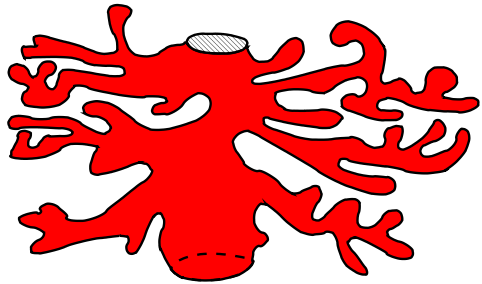}}}}
\vspace{-.6cm}
\caption[baby]{The fractal baby-universe structure of 2d
Euclidean gravity, artist's impression.}
\label{baby}
\end{figure}

A rough way of characterizing the quantum geometry is through its
Hausdorff dimension $d_{H}$. It can be measured by finding the 
scaling behaviour of the volumes $V(R)$ of geodesic balls of
radius $R$ in the ensemble of Lorentzian geometries,
\begin{equation}
\langle V(R)\rangle \sim R^{d_{H}}.
\end{equation}
It is straightforward to extract $d_{H}$ from the propagator,
yielding $d_{H}=2$. This may not seem a surprising result, since
we started from an ensemble of two-dimensional triangulations.
However, it is by no means a foregone conclusion, since it is
a property of the entire quantum ensemble (which, as we have seen, 
is subject to large fluctuations). Besides, we already know an
example where this does not happen, namely, Euclidean (or
``Liouville'') quantum gravity in two dimensions, which has
$d_{H}=4$! In this case, it is an indication of the highly fractal
nature of the quantum geometry, which is completely dominated by
so-called baby universes (Fig.\ \ref{baby}). 
Such highly branched configurations 
cannot occur in the Lorentzian state sum, since they are not compatible 
with our causality conditions. It can be shown explicitly that
this is the central difference between the two formulations, 
causing them to lie in different universality classes. This observation
has also been used to relate the two models by a renormalization
procedure that amounts to an ``integrating out of baby universes''
\cite{ackl}. It demonstrates that the relation between the two
continuum theories is rather more complicated than a simple
analytic continuation $t\mapsto - it$.

Thus we see that the physics described by the Lorentzian quantum 
gravity model is completely different from that of the Euclidean one.
Obviously, since quantum gravity in two dimensions is an unphysical 
theory, there is not much to choose between the two theories; we cannot 
perform experiments to determine which of them is ``correct'',
nor is there any {\it a priori} preference for a particular metric
signature. However, let us for the moment assume that we {\it were} 
interested in
obtaining a theory of {\it Lorentzian} geometries (as arguably
is the case in $d=4$). One could then argue that it was
unnatural to single out a ``time'' in the purely Euclidean
theory, since the fractal geometries have no distinguished directions
anywhere. Although it is clearly possible to make an arbitrary choice
of a time parameter, this will typically result in constant-time
slices that are highly multiply connected and undergo constant 
topology changes. Again, this may not be of great concern in dimension
2, but if a similar behaviour was found in $d=4$, one would have to make 
sure that it did not lead to consequences in
contradiction with observations.

\subsection{Coupling 2d Lorentzian gravity to matter}

I will now briefly describe the properties
of Lorentzian gravity coupled to matter fields. The partition function
for the coupled model takes exactly the same form as in Euclidean DT,
but again with the sum taken over causal triangulations only. 
For an Ising model with nearest-neighbour interaction it is given 
(in the Euclidean sector) by
\begin{equation}
Z=\sum\limits_{N_{2}}\e^{-\lambda N_{2}}
\sum\limits_{{\rm causal\ }T_{N_{2}}}\sum\limits_{\{\sigma_{i}=\pm 1\} }
\e^{\frac{\beta_{m}}{2}\sum\limits_{<ij>}\sigma_{i}\sigma_{j}},
\label{ising}
\end{equation}
where the last sum is over all possible spin configurations of the
Ising model on the triangulation $T_{N_{2}}$. We are interested both
in the critical properties of the matter on this non-trivial
``background'' and in possible back reactions of the matter on
the geometry, since the latter is represented by a fluctuating 
ensemble. Our main reference point for such a system is Euclidean
Liouville gravity with an Ising model, where the critical matter exponents
(specific heat, magnetization, 
magnetic susceptibility) are changed from
their Onsager values on fixed, regular lattices ($\alpha=0$, $\beta=1/8$,
$\gamma=7/8$) to $\alpha=-1$, $\beta=1/2$ and $\gamma=2$ \cite{bk}. 

The Lorentzian model has not yet been solved exactly, but we have
performed both a high-temperature expansion and Monte Carlo
simulations to determine its critical behaviour. 
(The diagrammatic expansion used in the former has some non-standard
features, since the graph counting takes place in a fluctuating
ensemble of geometries.)
At the combined
critical point of the cosmological coupling $\lambda$ and
the matter coupling $\beta_{m}$, they consistently yield the
{\it Onsager} exponents for the Ising matter \cite{aal1}. 
This may be surprising at first,
since one could have been tempted to interpret the outcome
of the Euclidean system cited above as an indication that the critical
matter behaviour must necessarily change in the presence of a
fluctuating geometry. Here we have an example where this is
not the case. Another lesson is that we also may not draw the converse
conclusion, namely, that Onsager exponents for the matter necessarily
imply that the underlying geometry is fixed and flat. 
On the contrary, we have to conclude that these exponents are
rather ``robust'', and that the geometry has to be very distorted
in order to cause a change of the critical matter behaviour.
(Note that one could try to turn this into a method for determining
critical matter exponents: simply couple them to Lorentzian lattices.
For the case of the Ising model, we found a remarkably good
convergence of the diagrammatic expansion for the susceptibility
$\gamma$ \cite{aal1}.)
\begin{figure}[h]
\vspace{-1.0cm}
\centerline{\scalebox{0.4}{\rotatebox{0}
{\includegraphics{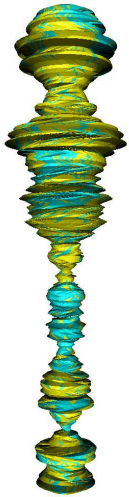}}}}
\vspace{-1.0cm}
\caption[ising1]{A typical Lorentzian geometry in the presence of one
Ising model, at volume $N_2=18816$ and total proper time $t=168$.}
\label{ising1}
\end{figure}

\subsection{... and more matter}

In 2d Lorentzian gravity coupled to
a single model of Ising spins, we did not find any appreciable
back reaction of the matter on the geometry (i.e. one that 
would have survived the continuum limit). However, as more matter
is coupled to the system ($n$ Ising models, with $n >1$), 
this is no longer true. The coupling is achieved by substituting
the last sum in (\ref{ising}) by a sum over $n$ independent copies
of the Ising model on the given triangulation $T_{N_2}$.
There is a very
good reason for studying this situation. In the language of conformal
systems, a system with $n$ Ising models at its critical point gives
rise to a conformal field theory with central charge $c=\frac{n}{2}$.
However, Liouville-matter models with $n>2$ are known to be inconsistent,
in the sense that their critical exponents become {\it complex}.
(This also goes by the name of ``$c=1$ barrier'' in bosonic string theory,
where $n$ plays the role of embedding dimension.)

\begin{figure}[h]
\vspace{-1.0cm}
\centerline{\scalebox{0.4}{\rotatebox{0}
{\includegraphics{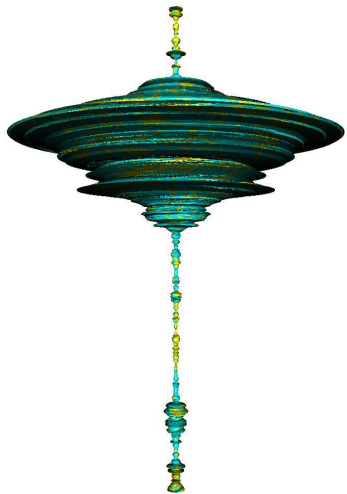}}}}
\vspace{-1.0cm}
\caption[ising8]{A typical Lorentzian geometry in the presence of eight
Ising models, at volume $N_2=73926$ and total proper time $t=333$.}
\label{ising8}
\end{figure}
In the Lorentzian case, we have found no such inconsistencies. We have
performed Monte Carlo simulations at $n=8$ \cite{aal2}, where the combined
system seems perfectly well-defined, and -- within the numerical error
bars -- the matter behaviour is again governed by the Onsager exponents! 
(The value 8 was chosen to be well beyond the region
$n=2$, since the experience from Euclidean dynamical triangulations
tells us that the phase change right at $n=2$ may not be very 
pronounced in numerical simulations.)
In contrast with the case $n=1$, we now observe a strong back reaction
on the geometry, which results in a {\it different universal behaviour} of
the gravitational sector. The impact of the matter coupling is best
illustrated visually. Fig.\ \ref{ising1} 
shows the coupling to a single Ising model
at $\lambda^{\rm crit}$ and $\beta_m^{\rm crit}$. As far as the geometry
of the configuration is concerned, there are no dramatic changes
compared with pure gravity (Fig.\ \ref{puregrav}). 
However, after switching on eight Ising models,
a typical configuration looks like Fig.\ \ref{ising8}. 
The effect of the matter
is to ``squeeze off'' part of the space-time to an effectively
one-dimensional region which will play no part in the continuum limit.
All interesting physics takes place in the remaining, extended part.
In this region of the geometry, we have measured the Hausdorff dimension
of space-time to be $d_H\approx 3$ \cite{aal2}. 
It is a tempting but completely
unproven {\it conjecture} that
the phase transition in the geometry takes place exactly at $c=1$.

We can understand the influence of the matter qualitatively, since 
the spin models have an energetic preference for {\it short}
boundaries between spin clusters of a given orientation. In a theory
where the geometry can fluctuate, spins will therefore have the
tendency to squeeze off part of the space-time geometry. In the
case of Euclidean Liouville gravity, whose geometries are very
branched to start with, this apparently leads to a complete
degeneracy of the geometry beyond the $c=1$ barrier. 
By contrast, the geometry of the Lorentzian 2d model remains 
well-defined.

\section{LORENTZIAN GRAVITY IN D=3} 
 
Which of the characteristic features of the
2d Lorentzian model generalize to higher dimensions and how
do they differ from their Euclidean counterparts? Our next stop
on the way to the physically relevant case $d=4$ is in three
dimensions. Apart from being a new statistical model of
three-dimensional fluctuating geometries, this theory has
some intrinsic interest. Although largely an unphysical
theory, 3d quantum gravity is an extensively studied system 
\cite{3dgrav}.
It is often invoked as a model system for the full theory,
since its classical equations resemble in many ways those
of general relativity. There are of course no physical
field degrees of freedom, and after getting rid of the 
diffeomorphisms, the theory has a finite-dimensional 
phase space. Although one has not yet been able to make full
use of this observation in a configuration space path-integral
formulation, it suggests that one may still be able to solve
3d gravitational models {\it analytically}. 

\begin{figure}[h]
\vspace{-.6cm}
\centerline{\scalebox{0.4}{\rotatebox{0}
{\includegraphics{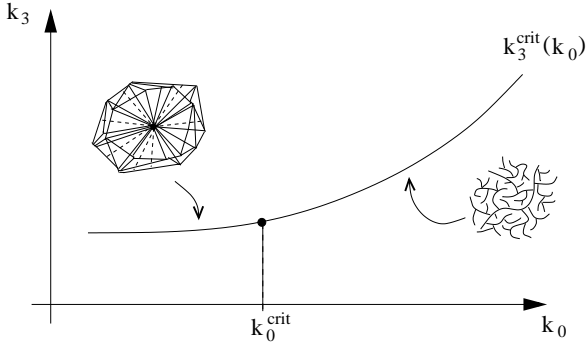}}}}
\vspace{-.6cm}
\caption[phaseeu]{The phase diagram of Euclidean dynamical 
triangulations in 3d.}
\label{phaseeu}
\end{figure}

As mentioned earlier, we have constructed an extension of the
simplicial Lorentzian formulation to 3 and 4 dimensions, on
a set of causal and ``Wick-rotatable'' geometries.
The model is
finite and well-defined at finite volumes, {\it without} the
need for further cut-offs \cite{ajl1,ajl2}. We have also shown that the
extreme geometric phases found in Euclidean dynamical triangulations
cannot be realized in the Lorentzian model. These phases of
rather degenerate geometry make up the phase diagram of
Euclidean DT in $d=3,4$ \cite{3deu,aj}, depicted in
Fig.\ \ref{phaseeu}. 
At small inverse gravitational coupling
$k_0\sim\frac{1}{G}$ one finds a ``crumpled'' phase, dominated by
configurations of very large Hausdorff dimension $d\approx\infty$
(these are simplicial manifolds where roughly speaking 
any two vertices are a 
minimal distance apart). Above the first-order
transition at $k_0^{\rm crit}$, the system is in a branched-polymer
phase of highly branched geometries
(with a fractal dimension $d_H=2$). 
Unfortunately, neither of these phases seems to have
a ground state that resembles an extended geometry of dimension
$d\geq 3$. 

The absence of these degenerate geometries from Lorentzian DT is
an encouraging feature, but only a kinematic property, 
which does not necessarily prevent the
occurrence of (less extreme) pathologies. To understand whether
Lorentzian gravity does indeed solve some of the problems of 
the Euclidean approach, we need to investigate its
phase structure by either numerical simulations or explicit
analysis. This work is still in progress, and I will summarize 
our current understanding of the three-dimensional case. More
technical details were reported by Ambj\o rn in the parallel
session \cite{amb2}, and can also be found in \cite{ajl3}.
In addition, efforts are under way to produce an analytical
solution, by using matrix models methods \cite{matrix} and
a continuum treatment of the gravitational path integral
in proper-time gauge \cite{pi}.

\begin{figure}[h]
\vspace{-.6cm}
\centerline{\scalebox{0.3}{\rotatebox{0}
{\includegraphics{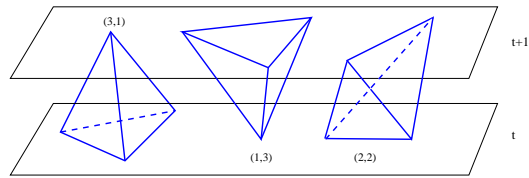}}}}
\vspace{-.6cm}
\caption[tetra]{The three types of tetrahedral building blocks
used in 3d Lorentzian gravity.}
\label{tetra}
\end{figure}

\subsection{Construction of 3d geometries}

The 3d Lorentzian space-times have again a foliated structure,
with spatial slices (of constant integer proper time $t$) 
given by equilateral Euclidean triangulations of the two-sphere.
The space in between these slices is filled by three types of
tetrahedral building blocks named (3,1), (1,3), and (2,2),
according to the number of vertices
they share with the spatial slices at times $t$ and $t$+1 
(Fig.\ \ref{tetra}).
The analogue of the 1+1 dimensional strips in Fig.\ \ref{2dtriang} 
are now 2+1 dimensional ``sandwiches'' $[t,t+1]$. 
As in 2d, the spatial edges have squared lengths $l_s^2=+a^2$,
and the time-like edges, interpolating between spatial slices,
have $l_t^2=-\alpha a^2$. We can then compute Regge's discretized
action $S_\alpha$, $\alpha >0$, 
for a given Lorentzian simplicial manifold of this type.
Since in the dynamical triangulations approach, both curvature
(i.e. deficit angles) and volume come in discrete units, 
the action can be written as a function of two ``bulk variables''
$N_d$ (the numbers of d-dimensional simplices), and the
total length $t$ of the geometry in proper time, leading to
a partition function of the form 
\begin{equation}
Z_\alpha (k_0,k_3,t)=\sum\limits_{{\rm causal\ }T}
\e^{iS_\alpha (N_0(T),N_3(T),t(T))}.
\label{3part}
\end{equation}
By a suitable analytic continuation in the complex $\alpha$-plane,
one finds that the Lorentzian and Euclidean actions are 
(for $\alpha <-1/2$, to satisfy Euclidean triangle inequalities) related by
\begin{equation}
S_{\alpha}(N_0,N_3,t)=S^{\rm Eu}_{-\alpha} (N_0,N_3,t). 
\end{equation}
For $\alpha =-1$, the right-hand side takes the standard form 
familiar from Euclidean DT,
\begin{equation}
S_1^{\rm Eu}= k_3N_3-k_0N_0,
\end{equation}
with the bare couplings ($\kappa\pi\equiv \arccos (1/3)$)
\begin{equation}
k_0=\frac{a}{4G},\ k_3=\frac{a}{4 G}(3 \kappa -
1)+\frac{a^3\Lambda}{48\sqrt{2}\pi G}.
\end{equation}

\begin{figure}[h]
\vspace{-.6cm}
\centerline{\scalebox{0.4}{\rotatebox{0}
{\includegraphics{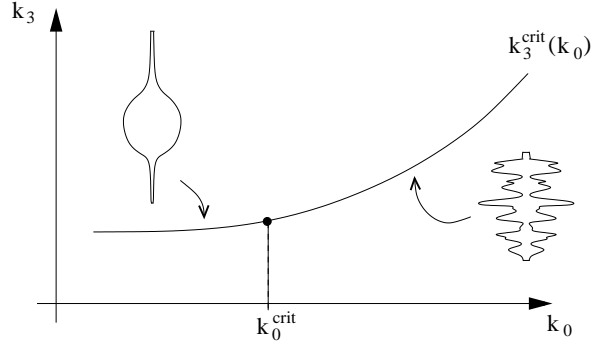}}}}
\vspace{-.6cm}
\caption[phaselor]{The phase diagram of Lorentzian dynamical
triangulations in 3d.}
\label{phaselor}
\end{figure}

\subsection{Phase structure of 3d Lorentzian gravity}

\begin{figure}[h]
\vspace{-2.8cm}
\centerline{\scalebox{0.4}{\rotatebox{0}{\includegraphics{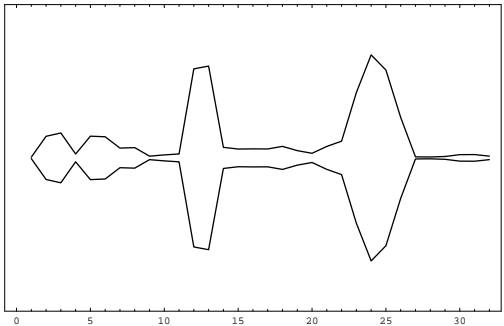}}}}
\vspace{-3.3cm}
\caption[ragged]{Snapshot of the distribution of two-volumes 
$N_{2s}(t)$ of spatial slices at proper times $t\in [0,32]$,
above the critical point $k_0^{\rm crit}$.}
\label{ragged}
\end{figure}

In order to study the 3d Lorentzian gravity path integral, we have set
up a Monte Carlo simulation for the Wick-rotated system with
partition function (\ref{3part}), at the Euclidean point $\alpha =-1$.
We have chosen $S^1\times S^2$ as a
convenient topology for our numerical purposes. The phase structure
can be characterized as follows (Fig.\ \ref{phaselor}). 
Just as in the Euclidean case,
we find a first-order transition point $k_0^{\rm crit}$ along the
critical line of the cosmological constant $k_3^{\rm crit}(k_0)$
(along which a continuum limit exists). However, the geometry of
the phases above and below this point 
are rather different. Above $k_0^{\rm crit}$,
the number of (2,2)-tetrahedra falls to a minimum, reducing the
space-time to an uncorrelated ensemble of successive spatial
slices, each well-described by Euclidean 2d gravity. This phase
does not seem interesting from a physical point of view, because
there are no long-range correlations in time-direction. 
A typical
configuration from this ``ragged'' phase is shown in 
Fig.\ \ref{ragged}.

A similarly degenerate phase, where the numbers of (1,3)- and 
(3,1)-tetrahedra attain a minimum, may exist for small $k_0$.
Since our algorithm is not efficient in this region,
we have not been able to explore whether there is a second
(possibly negative) critical point $\tilde k_0^{\rm crit}$. 
Wherever this second critical point may be, there is a remarkable
structure that emerges in the region of intermediate coupling,
$\tilde k_0^{\rm crit} <k_0 <k_0^{\rm crit}$, where all types of
tetrahedra contribute non-trivially. 
\begin{figure}[h]
\vspace{-2.2cm}
\centerline{\scalebox{0.4}{\rotatebox{0}{\includegraphics{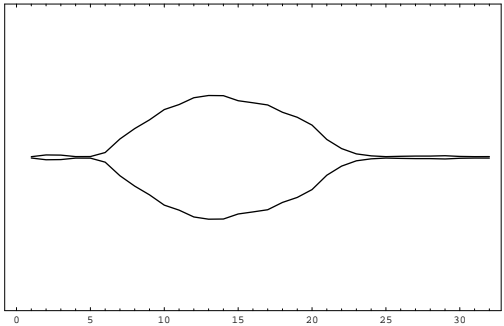}}}}
\vspace{-3.3cm}
\caption[universe]{Snapshot of the distribution of two-volumes 
$N_{2s}(t)$ of spatial slices at proper times $t\in [0,32]$,
below the critical point $k_0^{\rm crit}$.}
\label{universe}
\end{figure} 
\subsection{Geometry of the ground state in 3d}

In this intermediate phase, we observe the emergence of an 
extended geometry of roughly
spherical shape (Fig.\ \ref{universe} is a Monte Carlo ``snapshot''), and 
a definite extension $t_u$ in $t$-direction. This persists
in the simulations at all volumes $N_3$, provided the
total length $t$ of the compactified time-direction is chosen
sufficiently large, so that the ``universe'' can fit in.
It is surprising (and unprecedented) that such a structure should emerge
as the ground state of the quantum theory, given that we 
never put in any particular background geometry by hand.
It is also rather straightforward to see 
that its presence cannot be explained as a minimum of the Euclidean
action. It must therefore be the ground state of an {\it effective} action,
where entropy contributions (in other words, the {\it measure})
play a crucial role. Apparently in our model these contributions are
such that they outbalance potential conformal divergences coming
from the Euclidean action (otherwise a well-defined ground state
could not exist).

We have so far only investigated the macroscopic geometry
of the universe, i.e. its scaling properties at ``cosmic'' scales
\cite{ajl3}.
These are compatible with the scaling of a three-dimensional object.
Namely, the time extension $t_u$ scales according to $t_u\sim N_3^{1/3}$,
and the average two-volume of the spatial slices according to
$N_{2s}\sim N_3^{2/3}$. A final question concerns the role of
$k_0$ in this extended phase. The correlators between spatial volumes
and certain distributions of spherical disc volumes within given 
spatial slices
depend on $k_0$, but we have found that they can be mapped onto
each other by suitable (and equal) $k_0$-dependent rescalings of 
the lengths of time- and space-like links. This leads us to conclude
that the value of $k_0$ merely fixes an overall length scale,
and otherwise does not affect the physics of the model.

\section{SUMMARY}

I have given a brief overview of the current status of 
covariant lattice approaches to four-dimensional 
quantum gravity. Activity in the
area of Euclidean dynamical triangulations had somewhat slowed after 
a number of negative results concerning the nature of the phase
transition (although it has by no means been shown that gravity
{\it cannot} be quantized this way, if the current models are suitably
modified). However, even if this approach leads eventually
to a non-trivial continuum theory, some kind of ``Wick rotation''
will still be needed to make contact with physical geometries
of Lorentzian signature and with physical observables. 

A new class of {\it Lorentzian} dynamically triangulated models
presents an alternative to these Euclidean approaches.
Their starting point is a state sum over 
simplicial Lorentzian geometries, such that the Lorentzian 
nature of space-time
is built in from the outset. All of them have a distinguished
proper time, a well-defined causal structure, and can be 
uniquely Wick-rotated to Euclidean geometries. Topology changes of
the spatial slices are not allowed. 

From the point of view of statistical mechanics, 
they form a new class of models of random geometry (with a distinguished
direction or ``time arrow''). They are well-defined for finite
space-time volumes, in the sense that their transfer matrices are
bounded and strictly positive in dimension $d=2,3$ and 4, which implies the
existence of a self-adjoint Hamiltonian with a spectrum that is 
bounded from below.

We have found that in two and three dimensions, the properties of
the associated Lorentzian continuum theories 
are completely different from their
Euclidean counterparts. It seems that the causality conditions
imposed in the Lorentzian case act as an effective
``regulator'' on the geometry, still allowing for large local curvature
fluctuations, but suppressing the fractal structures found in
the Euclidean models in all dimensions. One consequence in two
dimensions was that we could cross without problems the infamous
$c=1$ barrier. In dimension three we observed, rather remarkably,
the emergence of a ground state of extended three-dimensional
geometry. 

These results are very encouraging. Motivated entirely by
physical considerations, we have discovered a new class of
dynamically triangulated models for quantum gravity. 
In $d=2,3$ we have found a number of new and interesting results.
We are particularly encouraged by the fact that the
three-dimensional Lorentzian model has a phase 
with a ground state of extended and non-degenerate geometry, because this
is exactly the point where the Euclidean DT model failed.
The ultimate test is of course gravity in four space-time
dimensions, where we expect a completely different picture,
with propagating field degrees of freedom
coming to the fore. 
It remains to be seen whether discrete Lorentzian quantum gravity
can bring us any closer to this holy grail ...

\vspace{.3cm} 
\noindent {\bf Acknowledgement.} 
I thank J.\ Ambj\o rn, K.N.\ Anagnostopoulos and
J.\ Jurkiewicz for enjoyable collaborations and A.\ Dasgupta and
D.\ Marolf for (equally enjoyable) discussions.

\end{document}